\documentclass[MSNbibl,nameyear,dvips]{arxstspdf}
\usepackage{flushend}
\usepackage{stfloats}

\volume{26}
\issue{1}
\pubyear{2011}
\firstpage{57}
\lastpage{58}
\doi{10.1214/11-STS345D}
\referstodoi{10.1214/10-STS345}

\makeatletter
\newcommand{\by}{\mathbf{y}}
\newcommand{\balpha}{\bolds{\alpha}}
\newcommand{\bGamma}{\bolds{\Gamma}}
\newcommand{\btheta}{\bolds{\theta}}
\newcommand{\wh}{\widehat}
\newcommand{\wt}{\widetilde}
\makeatother

\begin{document}
\begin{frontmatter}

\vspace*{12pt}
\title{Discussion of ``Feature Matching in Time Series Modeling'' by Y. Xia and H. Tong}
\runtitle{Discussion}
\pdftitle{Discussion of Feature Matching in Time Series Modeling by Y. Xia and H. Tong}

\begin{aug}
\author{\fnms{Qiwei} \snm{Yao}\corref{}\ead[label=e1]{q.yao@lse.ac.uk}}
\runauthor{Q. Yao}

\affiliation{London School of Economics}

\address{Qiwei Yao is Professor, Department of Statistics, London
School of
Economics, Houghton Street, London WC2A 2AE, United Kingdom \printead{e1}.}

\end{aug}


\vspace*{6pt}
\end{frontmatter}

Many congratulations to Professors Xia and Tong for another stimulating
paper initiated from
their own creative thinking.
The base point of the proposed approach is the fact that most, if not
all, statistical
models are wrong. This not only applies to time series models, as a
statistical model
is, hopefully, a simplified representation of the truth. At the best it
catches some features of
the unknown underlying population. While the understanding of this
nature is within the
common wisdom, most statistical inference methods are confined to the
framework which
assumes that the true model is a member of the family of models
concerned. The approach
advocated in this paper acknowledges explicitly that the assumed model
is not the truth, and
indeed it is advantageous sometimes not to read too much into the
assumed model.
For example, the authors have articulated elegantly that if our
interest lies in
catching the linear dynamical structure, we should not use the (Gaussian)
maximum likelihood estimation which effectively minimizes the one-step-ahead
prediction errors only, and in fact a better fitted autocovariance is
resulted from
minimizing up to $m$-step-ahead predictions for $m > 1$.

Following the lead of the authors, it seems to make sense to take on board
the concern for ``wrong models'' at the stage of the model selection, too,
as hinted at the end of the paper.
In the way, this has been actively researched in the context of model selection.
However, a~difference here is to use a different measure for ``goodness
of fit'' instead of likelihood
(or log-likelihood). Let us consider a simple case: fit a~linear $\operatorname{AR}(p)$
model to observations $y_1, \ldots, y_n$ from a~stationary time series
with mean
0, where the order $p$ is to be determined by the data, too. Let
$\by_{t,p} = (y_t, y_{t-1}, \ldots, y_{t-p+1})'$.
Based on an $\operatorname{AR}(p)$ model (with independent innovations), the best
predictor at the time $t$ for a future value $y_{t+m}$
should be a linear combination of the $p$ components of $\by_{t,p}$.
In fact the best linear predictor based on $\by_{t,p}$ is $\balpha
_{m, p}' \by_{t,p}$ with
\begin{eqnarray} \label{a1}
\balpha_{m, p} &=& \bGamma_p^{-1} \bolds{\gamma}_{m,p}\nonumber\\ [-8pt]\\ [-8pt]
 &=& \operatorname{arg\,min}\limits_{\bolds{\gamma}} E \{
(y_{t+m} - \balpha' \by_{t,p})^2\},\nonumber
\end{eqnarray}
where $\bGamma_p$ is a $p\times p$ matrix
with $\gamma(j-i)$ as its $(i,j)$th element,
$\bolds{\gamma}_{m,p}$ is a $p\times1$ vector with $\gamma(m+i
-1)$ as its $i$th element,
and $\gamma(\cdot)$ denotes the autocovariance
function of $y_t$.
In fact (1) holds for any stationary process. However, if we fit $y_t$
with an $\operatorname{AR}(p)$, its autocovariance function $\gamma(\cdot)$ is then
determined by $\btheta_p$---the parameters in an
 $\operatorname{AR}(p)$ model. Put $\balpha_{m, p} = \balpha_{m, p} (\btheta_p)$.
Then $\by_{t,p}'\balpha_{m, p} (\btheta_p)$ is the best predictor
for $y_{t+m}$ based on
an  $\operatorname{AR}(p)$ model.
Using the ``matching up-to-$m$-step-ahead point predictions'' approach of
Section 2.1, we estimate
$\btheta_p$ (for~$p$ given) by
\[
\wh\btheta_p = \operatorname{arg\,min}\limits_{\btheta_p} Q_p(\btheta_p),
\]
where
\begin{eqnarray*}
Q_p(\btheta_p) &=&
\frac{1}{m} \sum_{k=1}^m \frac{1}{n-k-p+1}\\
&&\hphantom{\frac{1}{m} \sum_{k=1}^m}{}\cdot \sum_{t=p}^{n-k} \{
y_{t+k} -
\by_{t,p}' \balpha_{k, p} (\btheta_p)\}^2.
\end{eqnarray*}
However, we cannot choose $p$ by minimizing $Q_p(\wh\btheta_p)$, as
$Q_p(\wh\btheta_p)$ is likely
to decrease as $p$ increases.

To appreciate the difficulties involved, let us first consider
the ``ideal world'' where the (true) distribution of $\{ y_t\}$ is known. Then
we should estimate~$\btheta_p$ by
\[\label{a3}
\wt\btheta_p = \operatorname{arg\,min}\limits_{\btheta_p} Q^*_p(\btheta_p),
\]
where
\[
Q_p^*(\btheta_p) =
{1 \over m} \sum_{k=1}^m E[\{ y_{t+k} -
\by_{t,p}' \balpha_{k, p} (\btheta_p)\}^2].
\]
Unfortunately $Q^*_p(\wt\btheta_p)$ still decreases as $p$ increases.
The information
(e.g., the variance) of the noise component of $y_t$ is required in
order to know when
to stop. This is the standard problem in model selection even for
linear regression.
One way to get away from this requirement is to take the
log-transforma\-tion. Namely, we define
%
\begin{eqnarray*}
L^*(p) &=& \log\{ Q_p^*(\wt\btheta_p)\}\\
 &=& \log\Biggl\{ \frac{1}{m} \sum_{k=1}^m E[\{ y_{t+k} -
\by_{t,p}' \balpha_{k, p} (\wt\btheta_p)\}^2] \Biggr\}.
\end{eqnarray*}
When $p$ is in the range on which $Q^*_p(\wt\btheta_p)$ varies slowly
(with respect to $p$), it holds that
\[
L^*(p) - L^*(p+1) \approx
\frac{ Q_p^*(\wt\btheta_p) - Q_{p+1}^*(\wt
\btheta_{p+1})}{Q_{p+1}^*(\wt\btheta_{p+1})}.
\]
Intuitively we would like to choose the smallest $p$ such that
the decrease $L^*(p) - L^*(p+1)$ is smaller than an appropriate but
unknown constant.
In practice, we may use $L(p) \equiv\log\{ Q_p(\wh\btheta_p)\} $ to
replace $L^*(p)$, and choose $p$ to minimize
\[
L(p) + E\{ L^*(p) - L(p) \}.
\]
This is in the same spirit of AIC in the sense that the bias $E\{
L^*(p) - L(p) \}$ serves
as the penalty for the model complexity. When the true model of $y_t$
is not\ $\operatorname{AR}$, this bias
does not admit a simple asymptotic expression such as AIC even when
$m=1$; see, for example, Konishi and
Kitagawa (\citeyear{KonKit96}). One may also consider to develop some resampling
estimates for this bias.

The above line of thinking is provoked from reading this interesting
paper which will serve
as an inspiration for further research in tackling the issues related
to the lack of a true model.
Then one may quibble over the use of the phrase ``catch-all approach.''
If a model
could catch all the features, it should be the true model, or at least
pragmatically so.
One message from the paper is that one should fit (and perhaps also
choose) a model
according to a specified purpose in hand, and a good statistical
modeling is to catch the
features of interest for a~particular purpose.

%

\end{document}